\documentclass[pra,twocolumn]{revtex4-2}
\usepackage{eurosym}
\usepackage{amssymb}
\usepackage{amsfonts}
\usepackage{amsmath}
\usepackage{graphicx}
\usepackage{bm}
\usepackage{braket}
\usepackage{epstopdf}
\usepackage{xcolor}

\begin{document}

\title{Stable hopfions in trapped quantum droplets}
\author{Zibin Zhao$^{1}$}
\thanks{These authors contributed equally to this work.}
\author{Guilong Li$^{1}$}
\thanks{These authors contributed equally to this work.}
\author{Huanbo Luo$^{1,2}$}
\email{huanboluo@fosu.edu.cn}
\author{Bin Liu$^{1,3}$}
\author{Gui-hua Chen$^4$}
\email{cghphys@gmail.com}
\author{Boris A. Malomed$^{5,6}$}
\author{Yongyao Li$^{1,3}$ }
\email{yongyaoli@gmail.com}
\affiliation{$^1$School of Physics and Optoelectronic Engineering, Foshan University,
Foshan 528000, China\\
$^2$Department of Physics, South China University of Technology, Guangzhou
510640, China\\
$^3$Guangdong-Hong Kong-Macao Joint Laboratory for Intelligent Micro-Nano
Optoelectronic Technology, Foshan University, Foshan 528225, China\\
$^{4}$School of Electronic Engineering \& Intelligentization, Dongguan
University of Technology, Dongguan 523808, China\\
$^5$Department of Physical Electronics, School of Electrical Engineering,
Faculty of Engineering, Tel Aviv University, Tel Aviv 69978, Israel\\
$^6$Instituto de Alta Investigaci\'{o}n, Universidad de Tarapac\'{a},
Casilla 7D, Arica, Chile }

\title{Stable hopfions in trapped quantum droplets}

\begin{abstract}
Hopfions are a class of three-dimensional (3D) solitons which are built as
vortex tori carrying intrinsic twist of the toroidal core. They are
characterized by two independent topological charges, \textit{viz}.,
vorticity $S$ and winding number $M$ of the intrinsic twist, whose product
determines the \textit{Hopf number}, $Q_{H}=MS$, which is the basic
characteristic of the hopfions. We construct hopfions as solutions of the 3D
Gross-Pitaevskii equations (GPEs) for Bose-Einstein condensates in binary
atomic gases. The GPE system includes the cubic mean-field self-attraction,
competing with the quartic self-repulsive Lee-Huang-Yang (LHY) term, which
represents effects of quantum fluctuations around the mean-field state, and
a trapping toroidal potential (TP). A systematic numerical analysis
demonstrates that families of the states with $S=1,M=0$, i.e., $Q_{H}=0$,
are stable, provided that the inner TP\ radius $R_{0}$ exceeds a critical
value. Furthermore, true hopfions with $S=1,M=1\sim 7$, which correspond,
accordingly, to $Q_{H}=1\sim 7$, also form partly stable families, including
the case of the LHY\ superfluid, in which the nonlinearity is represented
solely by the LHY term. On the other hand, the hopfion family is completely
unstable in the absence of the LHY term, when only the mean-field
nonlinearity is present. We illustrate the knot-like structure of the
hopfions by means of an elementary geometric picture. For $Q_{H}=0$, circles
which represent the \textit{preimage} of the full state do not intersect. On
the contrary, for $Q_{H}\geq 1$ they intersect at points whose number is
identical to $Q_{H}$. The intersecting curves form multi-petal structures
with the number of petals also equal to $Q_{H}$.
\end{abstract}

\maketitle


\section{Introduction}

In 1867, Lord Kelvin proposed the hypothesis of vortex atoms \cite%
{Kelvinvortex-atom}, sparking mathematical studies of knots. However, it was
not until 1997 that Faddeev and Niemi suggested that knots might exist as
stable solitons in the three-dimensional classical field theory \cite%
{Faddeevknotnature}, thereby initiating research of physical realizations of
knot-like structures. Numerical studies of the Faddeev-Skyrme version of the
O(3) sigma model have demonstrated the existence of stable toroidal
solutions in it \cite%
{GladikowskiHopf1997,Battyeknots1998,Aratyn3+1integrable,Hietarinta1999PLB,Aratynprl1999, RSWardHopfsolitons,sublinearknot2007}%
. Unlike ordinary vortex-torus solutions, which are characterized by the
single winding number (topological charge) $M$, those ones possess two
independent winding numbers, $M$ and $S$, with $S$ determining the \textit{%
inner twist} of the vortex torus (see Eqs. (\ref{input}) and (\ref%
{coordinates}) below). The corresponding Hopf number is defined as
\begin{equation}
Q_{H}=MS,  \label{QMS}
\end{equation}%
which is the fundamental topological characteristic of the intrinsically
twisted vortex-soliton states, known as \textit{hopfions}.

Hopfions, which constitute a class of knot-like solitons, may be elegantly
presented by dint of the \textit{Hopf fibration}, which is a mapping from
the unit sphere in the four-dimensional space onto unit sphere in 3D, i.e., $%
S^{3}\rightarrow S^{2}$ \cite{RaduknotPR2008}. Various species of hopfions
have been extensively studied across many fields \cite{Gobelbeyondskrmions}-%
\cite{FNRybakovmagnetic}.

In optics, scalar hopfions with toroidal vortex structures have been
proposed as approximate solutions to the Maxwell's equations \cite%
{Qiwenzhanelight}. Additionally, the use of structured light fields,
featuring pronounced spatial variations in their polarization, phase, and
amplitude, have made it possible to demonstrate an experimental realization
of optical hopfions \cite{YjshenAP2023}. Recent studies employing high-order
harmonics have also produced optical hopfions \cite{ZlyuHHGhopfions}. In
liquid crystals, material structures in the form of hopfions and various
bound states built of them have been widely investigated too, both
theoretically and experimentally \cite%
{Ackermanhopfionarticles,Smalyukhreview,jsbtai3Dsrystals,JSBTaiTopologicaltransformations,HROSohnDynamics,Ivan}%
. In cosmology, topological defects are utilized to model the large-scale
structure of the universe, with studies of knot-like hopfions providing
valuable insights into the topology of cosmic formations \cite%
{vilenkincosmic,Hojmancosmichopfions,fostermassive,Thompsonalgebraichopfions, smolkagravitationalhopfions}%
. These findings highlight the significance of hopfions in unraveling
complex topological phenomena across diverse disciplines.

Ultracold atomic Bose-Einstein condensates (BECs), with their tunable
intrinsic interactions, offer an expedient platform for the realization of
knot-shaped solitons. Various types of vortex knot solitons have been
extensively studied in BECs, including trefoil vortex knots, Solomon vortex
links, and various other vortex rings and lines \cite%
{promentvortexknot,hbluo3Dskyrmions,
wkbaivortexring,Uedaknotsspinor,DShall,rnbissetlineringhopfions,
Ollikainenquantumknots,YKLiuknotspin1, BJackson99pra,WWangvortexring,
CTicknorspectra,VARubanInstabilities,MAbadvortexring}. Topological
transitions from vortex rings to knots and links were explored too \cite%
{wkbaitopologicaltransition}. In particular, encircling a vortex line by a
vortex ring gives rise to composite hopfions \cite{Bidasyukhopfsoliton}. 
Previous studies have demonstrated that
hopfions can stably exist in a rotating BEC confined by an oblate
harmonic-oscillator trap \cite{Bidasyukhopfsoliton}. Moreover, in the 3D
free space, hopfions may be stable in a single-component BEC with
self-repulsion whose strength is made spatially modulated, growing, as a
function of radial coordinate $r$, faster than $r^{3}$ \cite%
{Kartashovtoroidalvortex}.

Quantum droplets (QDs), i.e., self-trapped states filled by an ultradilute
superfluid, are maintained by the balance between mean-field effects and
corrections to them induced by quantum fluctuations \cite%
{PetrovLHYterm15,PetrovLHYterm16}. In terms of the corresponding
Gross-Pitaevskii equations (GPEs), the corrections are represented by the
Lee-Huang-Yang (LHY) terms \cite{jorgensenQuantumfluctation,leehardsphere}.
Actually, QDs represent a new species of quantum matter, as well as a novel
form of self-localized states \cite{ZHluonewformliquid,TPfaunewstate}.
Recent detailed studies predict that stable QDs may exist, in the free
space, not only in the form of a self-trapped ground-state (GS) mode, but
also as robust excited states with embedded vorticity \cite%
{LYY2DQDs,LGLDDIQDs,XXXprl,Kartashov3DQDs,DlwrotatemultDQDs,ZLingettink,
TAYogurtvortexlattice,BAMalomeddynamics1Dqds,CDErricoobserqds2019,TPFauobserindipolar,rnbissetQDsdipolar}%
. In addition to that, stable vortex QDs with high values of the topological
charge (vorticity), as well as multipole QDs, can be maintained by a
Gaussian-shaped toroidal potential (TP)\ \cite{Dlw3Dqdstoroidal}. However,
three-dimensional (3D) vortex states with complex topological structures,
such as hopfions, have yet to be investigated in the realm of QDs.

In this paper, we aim to produce stable solutions for hopfion QDs built with
the help of TP. Unlike the above-mentioned one trapping potential, which was
employed in Ref. \cite{Dlw3Dqdstoroidal}, the TP considered here is one of
the harmonic-oscillator type, rather than Gaussian-shaped potential
structure. Numerical analysis indicates that the hopfions with $Q_{H}=0$
(they may exist with $M=0$ and $S\neq 0$, see Eq. (\ref{QMS}) and Ref. \cite%
{Kartashov3DQDs}, as toroidal patterns whose \textit{preimages}
(\textquotedblleft skeletons") \cite{Smalyukhreview} are composed of
non-intersecting concentric circles) are stable, in a typical experimentally
relevant setup, at values of the TP's inner radius $R_{0}\geq 1.1\,\mathrm{%
\mu }\text{m}$, an instability region appearing below this threshold.
Furthermore, hopfions with high values of the Hopf number, up to $Q_{H}=7$,
are also found to be stable. In particular, in contrast to the results of
Ref. \cite{Bidasyukhopfsoliton}, the inclusion of the LHY term allows for
the existence of stable hopfions with $S=1$ and $M>1$ (hence they have $%
Q_{H}>1$, as per Eq. (\ref{QMS})). Similar to the $Q_{H}=0$ solutions, these
hopfions exhibit a double-ring pattern, which may serve as an experimental
signature for identifying hopfions. These hopfion modes correspond to
\textit{preimages} which include intersections. In the horizontal plane, the
self-intersecting preimages form petal-like structures, see Fig. \ref{knots}
below. As $Q_{H}$ increases, the number of petals increases accordingly.
These findings provide direct insights into the geometric and topological
structures of hopfions in QDs and suggest a potential pathway for their
experimental realization.

The subsequent presentation is organized as follows. Section 2 introduces
the 3D model including the TP. Numerical results for stationary solutions,
including the hopfions with $Q_{H}=0$ and $Q_{H}\neq 0$, are reported in
Section 3. Additionally, an elementary geometric representation of the Hopf
number is introduced and analyzed in that section. The paper is concluded by
Section 4.

\section{The model}

The binary BEC in the 3D space with coordinates $\left( X,Y,Z\right) $ is
modeled by the system of nonlinearly-coupled GPEs which include the cubic
mean-field terms and quartic LHY ones \cite{PetrovLHYterm15}:

\begin{equation}
\begin{split}
i\hbar \frac{\partial }{\partial T}\Psi _{1}=& -\frac{\hbar ^{2}}{2m}\nabla
_{\mathrm{XYZ}}^{2}\Psi _{1}+\left( G_{11}|\Psi _{1}|^{2}+G_{12}|\Psi
_{2}|^{2}\right) \Psi _{1} \\
& +\Gamma \left( |\Psi _{1}|^{2}+|\Psi _{2}|^{2}\right) ^{\frac{3}{2}}\Psi
_{1}+V\left( R,Z\right) \Psi _{1},
\end{split}
\label{coupled-GPE1}
\end{equation}%
\begin{equation}
\begin{split}
i\hbar \frac{\partial }{\partial T}\Psi _{2}=& -\frac{\hbar ^{2}}{2m}\nabla
_{\mathrm{XYZ}}^{2}\Psi _{2}+\left( G_{22}|\Psi _{2}|^{2}+G_{21}|\Psi
_{1}|^{2}\right) \Psi _{2} \\
& +\Gamma \left( |\Psi _{1}|^{2}+|\Psi _{2}|^{2}\right) ^{\frac{3}{2}}\Psi
_{2}+V\left( R,Z\right) \Psi _{2},
\end{split}
\label{coupled-GPE2}
\end{equation}%
where $G_{11}=G_{22}=4\pi \hbar ^{2}a/m$ and $G_{12}=G_{21}=4\pi \hbar
^{2}a^{\prime }/m$ are the self- and cross-interaction strengths, with
atomic mass $m$, $a$ and $a^{\prime }$ being the intra- and inter-species
scattering lengths, respectively. The TP potential is defined as%
\begin{equation*}
V\left( R,Z\right) =\frac{1}{2}m\Omega ^{2}\left[ \left( R-R_{0}\right)
^{2}+Z^{2}\right] ,
\end{equation*}%
where $R=\sqrt{X^{2}+Y^{2}}$ is the radial coordinate in the 2D plane, $%
R_{0} $ is the inner radius of the toroid, and $\Omega $ is the TP trapping
frequency. The coefficient of the LHY correction is \cite{PetrovLHYterm15}%
\begin{equation}
\Gamma =\frac{4m^{\frac{3}{2}}G^{\frac{5}{2}}}{3\pi ^{2}\hbar ^{3}}=\frac{128%
\sqrt{\pi }}{3m}\hbar ^{2}a^{\frac{5}{2}}.  \label{LHY}
\end{equation}

For symmetric states with
\begin{equation}
\Psi _{1}=\Psi _{2}\equiv \Psi /\sqrt{2},  \label{symmetric-condition}
\end{equation}%
coupled GPEs (\ref{coupled-GPE1}) and (\ref{coupled-GPE2}) admit the
reduction to a single equation,
\begin{equation}
\begin{split}
i\hbar \frac{\partial }{\partial T}\Psi =& -\frac{\hbar ^{2}}{2m}\nabla _{%
\mathrm{XYZ}}^{2}\Psi +\frac{\delta G}{2}\left\vert \Psi \right\vert ^{2}\Psi
\\
& +\Gamma \left\vert \Psi \right\vert ^{3}\Psi +\frac{1}{2}m\Omega ^{2}\left[
\left( R-R_{0}\right) ^{2}+Z^{2}\right] \Psi ,
\end{split}
\label{GPE}
\end{equation}%
where $\delta G=\left( 4\pi \hbar ^{2}/m\right) \left( a^{\prime }+a\right)
\equiv \left( 4\pi \hbar ^{2}/m\right) \delta a$. The total number of atoms
in the system is
\begin{equation}
N=\int \left( \left\vert \Psi _{1}\right\vert ^{2}+\left\vert \Psi
_{2}\right\vert ^{2}\right) d^{3}\mathbf{R}=\int \left\vert \Psi \right\vert
^{2}d^{3}\mathbf{R}.  \label{atom-Number}
\end{equation}%
The set of the system's control parameters includes $a$, $a^{\prime }$ and $%
N $.

By means of rescaling,
\begin{equation}
T=t\cdot t_{0},\left( X,Y,Z\right) =\left( x,y,z\right) \cdot l_{0},\Psi
=l_{0}^{-\frac{3}{2}}\cdot \psi ,\frac{\hbar t_{0}}{ml_{0}^{2}}=1,
\label{rescaling}
\end{equation}%
where $t_{0}$ and $l_{0}$ are time and length scales, Eq. (\ref{GPE}) is
cast in the dimensionless form:
\begin{equation}
\begin{split}
i\frac{\partial }{\partial t}\psi =& -\frac{1}{2}\nabla ^{2}\psi
+g\left\vert \psi \right\vert ^{2}\psi +\gamma \left\vert \psi \right\vert
^{3}\psi \\
& +\frac{1}{2}\omega \left[ \left( r-r_{0}\right) ^{2}+z^{2}\right] \psi ,
\end{split}
\label{dimensionless-GPE}
\end{equation}%
where we define the scaled strengths of the contact interaction, LHY
correction, and TP as, respectively,
\begin{equation}
\begin{split}
& g=2\pi \frac{\delta a}{l_{0}}=2\pi \delta \tilde{a}, \\
& \gamma =\frac{128\sqrt{\pi }}{3}\left( \frac{a}{l_{0}}\right) ^{\frac{5}{2}%
}=\frac{128\sqrt{\pi }}{3}\tilde{a}^{\frac{5}{2}}, \\
& \omega=\left( t_{0}\Omega \right) ^{2}.
\end{split}
\label{dimensionless-parameters}
\end{equation}%
In the further analysis, we refer to parameters of the $^{39}\mathrm{K}$
atomic gas, selecting $l_{0}=0.1~\mathrm{\mu }$m. The Hamiltonian (energy)
corresponding to Eq. (\ref{dimensionless-GPE}) is
\begin{equation}
\begin{split}
E=\int \Big(& \frac{1}{2}\left\vert \nabla \psi \right\vert ^{2}+\frac{1}{2}%
g\left\vert \psi \right\vert ^{4}+\frac{2}{5}\gamma \left\vert \psi
\right\vert ^{5} \\
& +\frac{1}{2}\omega\left[ \left( r-r_{0}\right) ^{2}+z^{2}\right]
\left\vert \psi \right\vert ^{2}\Big)d^{3}\mathbf{r}.
\end{split}
\label{energy}
\end{equation}

Stationary solutions of Eq. (\ref{dimensionless-GPE}) with chemical
potential $\mu $ are looked for, below, as
\begin{equation}
\psi =\phi (r)\exp (-i\mu t),  \label{stationary-solutions}
\end{equation}%
with function $\phi $ obeying the stationary GPE:
\begin{equation}
\begin{split}
\mu \phi =& -\frac{1}{2}\nabla ^{2}\phi +g\left\vert \phi \right\vert
^{2}\phi +\gamma \left\vert \phi \right\vert ^{3}\phi \\
& +\frac{1}{2}\omega \left[ \left( r-r_{0}\right) ^{2}+z^{2}\right] \phi .
\end{split}
\label{stationary-GPE}
\end{equation}

\begin{figure}[tbp]
{\includegraphics[width=3.1in]{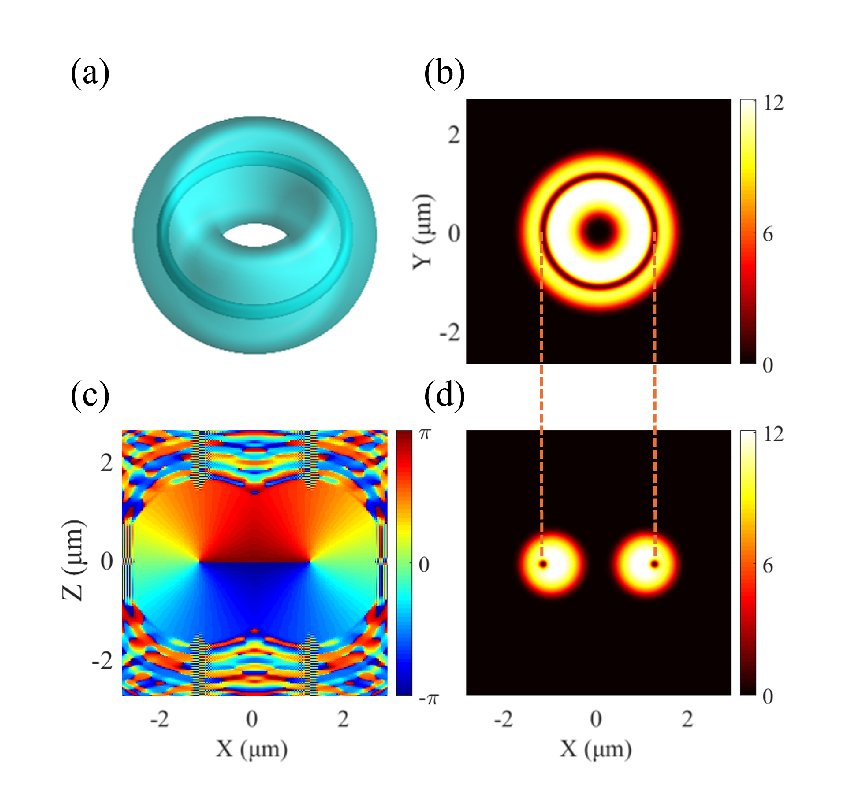}}
\caption{An example of a stable hopfion with vorticity $M=0$, inner twist $%
S=1$, trapping frequency $\Omega =51000\text{ Hz}$, atom number $N=55492$,
inner radius $R_{0}=0.9$ $\mathrm{\protect\mu }$m, $a=100a_{0}$ and $%
a^{\prime }=-110a_{0}$. The corresponding dimensionless parameters are $%
g=-0.03$, $\protect\gamma =0.05$, and $\protect\omega=0.1$. (a) A density
isosurface of the stable toroidal vortex. (b) The density distribution in
the horizontal plane, $Z=0$. (c-d): The phase and density distributions of
the wave function in the vertical plane, $Y=0$. Orange dashed lines
designate positions of pivots (phase singularities) of the internal vortex
rings, as observed in the $Y=0$ plane.}
\label{s=1-m=0}
\end{figure}

\section{Stationary solutions}

Hopfions are defined as knot-like solitons that can be mapped into the Hopf
fibration \cite{hopf1964}, being characterized by two independent winding
numbers \cite{lyonshopffibration,Whiteheadhopfinvariant}. The numerical
solution of Eq. (\ref{dimensionless-GPE}) for hopfions was performed by
means of the Newton-conjugate-gradient method, starting with the ansatz
(initial guess)%
\begin{equation}
\phi \left( r,z\right) =\left( r^{\prime }\right) ^{S}\exp \left( -\frac{%
\left( r^{\prime }\right) ^{2}}{A}+iM\theta +iS\varphi \right) ,
\label{input}
\end{equation}%
where $A>0$ is a real constant and the special coordinates are defined as
\begin{equation}
\left( r^{\prime },\theta ,\varphi \right) =\left[ \sqrt{\left(
r-r_{0}\right) ^{2}+z^{2}},\arctan \frac{y}{x},\arctan \frac{r-r_{0}}{z}%
\right]  \label{coordinates}
\end{equation}%

(i.e., $\theta $ is the usual angular coordinate in the horizontal plane $%
\left( x,y\right) $). In contrast with the usual vortex tori in the 3D
space, which feature the single winding number $M$ (vorticity) in the $(x,y)$
plane, here the second winding number $S$, which is defined in the $(r,z)$
plane, determines the inner twist of the vortex torus. Ansatz (\ref{input})
implies a structure like a\ twisted toroidal vortex tube nested in the 3D
solution, coiling up around the vertical ($z$) axis. This shape is typical
for solitons of the Faddeev-Skyrme model, with the triplet of real scalar
fields realizing the Hopf map, $\mathbf{\Phi }:R^{3}\rightarrow S^{2}$ \cite%
{Faddeevknotnature,FaddeevPRL1999,GladikowskiHopf1997,jaykkaPRB2008,babaevknotsoliton,speightSupercurrent,jaykkaPRD2011}%
, therefore such states are named hopfions, which are characterized by the
Hopf number (topological invariant) (Eq. \ref{QMS}).

\begin{figure}[tbp]
{\includegraphics[width=3.4in]{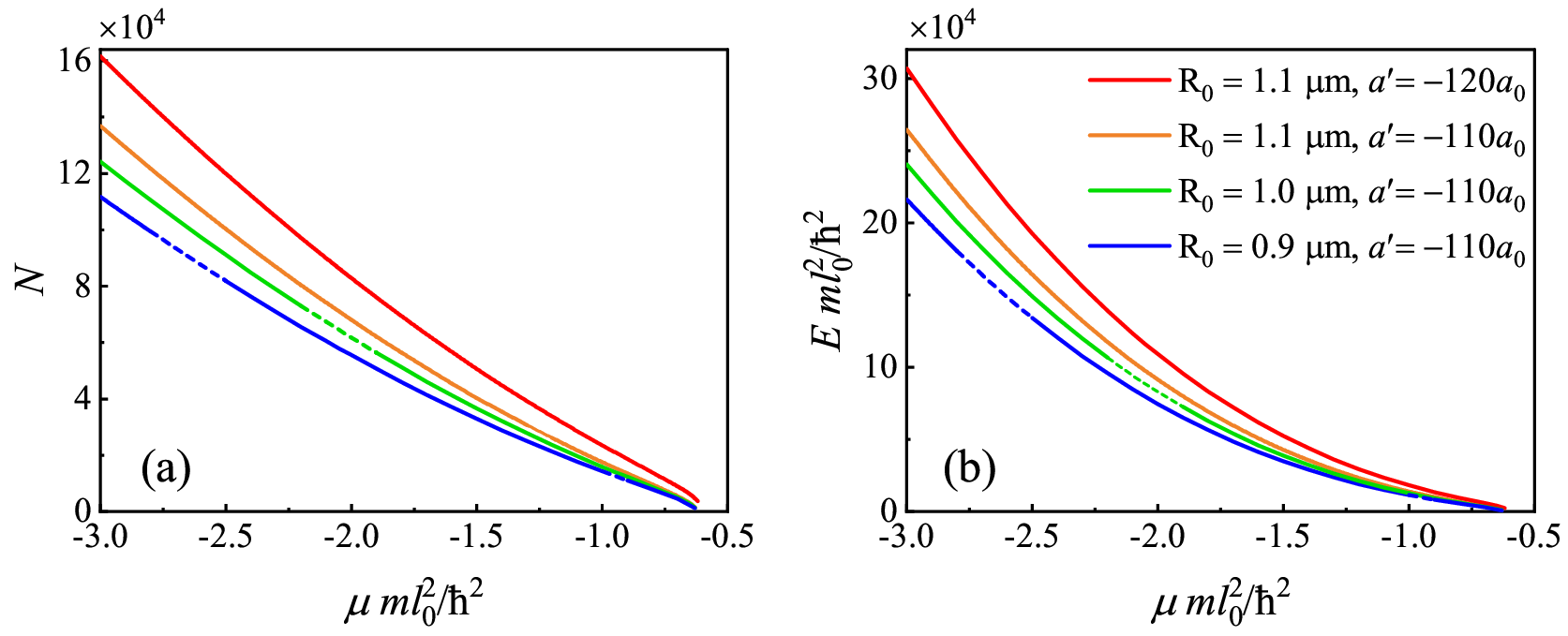}}
\caption{The total atom number $N$ and energy $E$ are plotted vs. the
chemical potential $\protect\mu $ for hopfions with $S=1$, $M=0$ in panels
(a) and (b), respectively. Blue and green lines represent hopfions with $%
a^{\prime }=-110$ $a_{0}$, $R_{0}=0.9$ $\mathrm{\protect\mu }$m and $%
R_{0}=1\,\mathrm{\protect\mu }$m, respectively, while the orange and red
lines correspond to $R_{0}=1.1$ $\mathrm{\protect\mu }$m, $a^{\prime }=-110$
$a_{0}$, and $a^{\prime }=-120$ $a_{0}$, respectively. Solid and dashed
segments indicate stable and unstable hopfions, respectively. In all cases,
we set $\Omega =51000\,\text{Hz}$, $a=100$ $a_{0}$, the corresponding
dimensionless parameters being $\protect\omega =0.1$ and $\protect\gamma %
=0.05$. When $a^{\prime }=-110$ $a_{0}$ and $-120~a_{0}$, the corresponding
dimensionless nonlinearity strength is $g=-0.03$ and $-0.06$, respectively.}
\label{NEmu}
\end{figure}

\subsection{Hopfions with $Q_{H}=0$}

A typical stable hopfion with zero vorticity, i.e., $S=1,M=0$ (hence it has $%
Q_{H}=0$, according to Eq. (\ref{QMS})), which is supported by the TP, is
displayed in Fig. \ref{s=1-m=0}. Direct simulations of its perturbed
evolution demonstrate that this hopfion is stable, at least, up to $T=10$
ms. Its toroidal shape produces a double-ring pattern in the horizontal
plane $Z=0$, as shown in Fig. \ref{s=1-m=0}(b), where density $|\phi |^{2}$
of the inner ring is slightly higher than in the outer ring. The position of
the pivots (phase singularities) of the internal vortex rings at $Y=0$,
corresponding to the location of the inner ring, is indicated by the orange
dashed line in Fig. \ref{s=1-m=0}. This double-ring pattern plays the role
similar to that of the hopfion's \textit{preimage} \cite{Smalyukhreview},
and may serve as a signature for the its experimental observation.

\begin{figure}[tbp]
{\includegraphics[width=3.4in]{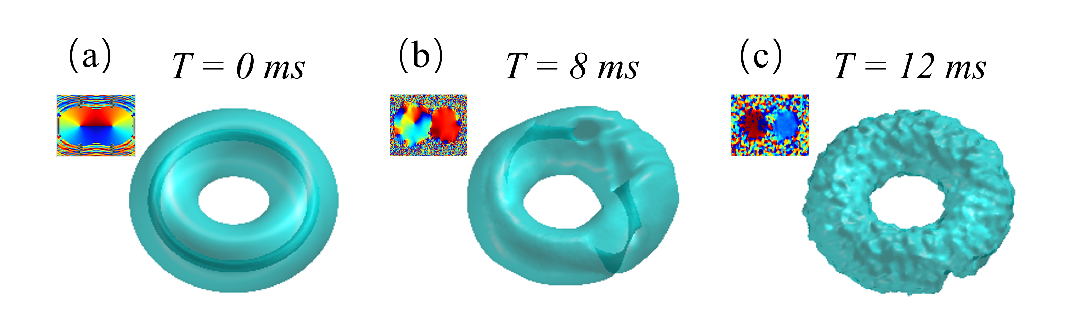}}
\caption{The unstable evolution of a hopfion with $S=1$, $M=0$, $R_{0}=0.9\,%
\mathrm{\protect\mu }$m, $a=100$ $a_{0}$, $a^{\prime }=-110$ $a_{0}$, and $%
N=32870$. Panels (a), (b), and (c) display the hopfions at $T=0\,\text{ms}$,
$8\,\text{ms,}$ and $T=12\,~\text{ms}$, respectively.}
\label{unstable-QH0}
\end{figure}

Basic properties of the hopfions with $S=1,M=0$ are summarized in Fig. \ref%
{NEmu}. It displays the total atom number $N(\mu )$ and energy $E(\mu )$ for
the hopfions with different values of $a^{\prime }$ and $R_{0}$, with solid
lines and dashed segments referring to stable and unstable solutions,
respectively. It is evident that, for a fixed scattering length $a$,
increase in the\ TP radius $R_{0}$ leads to higher atom number $N$ and
energy $E$. This\ fact suggests that a larger radius requires a higher
number of atoms for the formation of hopfions. Additionally, an instability
region arises when $R_{0}$ is below $1.1\,\mathrm{\mu }\text{m}$. This\
observation implies that, for a fixed interaction strength $g$, the hopfions
suffers destabilization when the inner TP radius $R_{0}$ falls below a
critical value, $R_{c}=1.1$ $\mathrm{\mu }\text{m}$. A more detailed investigation 
reveals that Hopfions with $S = 1, M = 0$ cease to exist when $R_0$ falls 
below $0.4\,\mu\text{m}$. This indicates that in the case of a spherically 
symmetric harmonic trap, corresponding to $R_0 = 0$, such Hopfion structures 
cannot exist in a stable form.

Note that all the $N(\mu )$ dependences plotted in Fig. \ref{NEmu}(a)
satisfy the \textit{Vakhitov-Kolokolov criterion}, $dN/d\mu <0$, which is
the well-known necessary stability criterion \cite{VK,Berge}. On the other
hand, it is seen too that the criterion is not sufficient for the full
stability, as the curves include unstable segments. The same conclusions are
valid for the $N(\mu )$ dependences for the hopfion families with $Q_{H}\neq
0$, which are plotted below in Fig. \ref{QHne0NEmu}.

The simulated evolution of an unstable hopfion with $S=1$, $M=0$ is shown in
Fig. \ref{unstable-QH0}. In the course of the evolution, it loses the phase
structure and eventually decay into the GS. Although Fig. \ref{unstable-QH0}
illustrates the evolution up to $T=12\,\text{ms}$, the ring-like structure
of weakly unstable hopfions can be preserved for a relatively long time and
does not split until $T=40\,\text{ms}$. Moreover, for the same values of $a$
and TP radius $R_{0}$, increase in $|a^{\prime }|$, which corresponds to
enhancement of the interaction strength $g$, results in an increase in both
the number of atoms $N$ and energy $E$. As we adopt $a^{\prime }<(-a)<0$, it
follows from here that we have $g<0$, indicating that the increase of $%
|a^{\prime }|$ strengthens the attractive interaction. 
Thus, stronger attraction leads to the formation of hopfions
with higher atom numbers and energy. Actually, the hopfions with $S=1,M=0$
exhibit the lowest atom-number threshold, below which the hopfions do not
exist. This threshold increases with the TP radius $R_{0}$ and interaction
strength $g$. 

\begin{figure}[tbp]
{\includegraphics[width=2.5in]{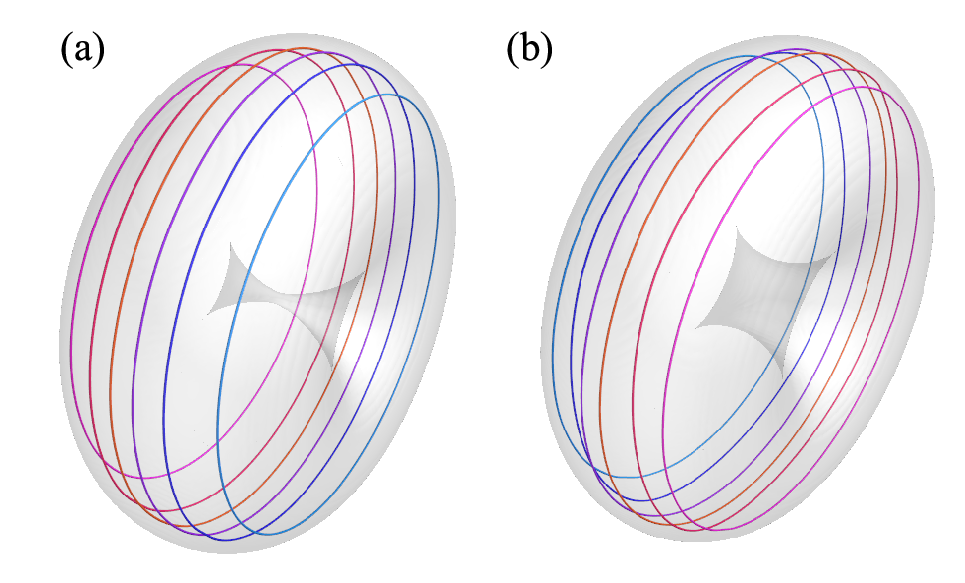}}
\caption{Isolines of the hopfions with $S=1,M=0$, $a=100~a_{0}$, $a^{\prime
}=-110~a_{0}$, $N=55492,$ $R_{0}=0.9$ $\mathrm{\protect\mu }$m (a) and $%
N=68006$, $R_{0}=1.1$ $\mathrm{\protect\mu }$m (b). Different colors
represent different constant values of $(\text{Re}(\protect\phi ),\text{Im}(%
\protect\phi ))$, which, from right to left, are: $(0.3,0.4)$, $(0.5,0.3)$, $%
(0.6,0.1)$, $(0.3,-0.4)$, $(0.5,-0.3)$, and $(0.6,-0.1)$. In (a) and (b),
the gray color represent the isosurface with $|\protect\phi |^{2}=0.4$.}
\label{equalvalueline}
\end{figure}

As mentioned above, hopfions are a class of solitons of the Faddeev-Skyrme
model and realize the Hopf map, $\mathbf{\Phi }:R^{3}\rightarrow S^{2}$. For
finite-energy solutions, one requires $\mathbf{\Phi }\rightarrow \mathbf{n}$
as $|\mathbf{r}|\rightarrow \infty $, where $\mathbf{n}$ is a constant unit
vector. Thus $R^{3}$ can be compactified to $S^{3}$ and the map reduces to
\begin{equation}
\mathbf{\Phi }:S^{3}\rightarrow S^{2}.  \label{Hopfmap}
\end{equation}%
Since $\pi _{3}\left( S^{2}\right) =\mathbb{Z}$, where different integers
from $\mathbb{Z}$ are the Hopf numbers $Q_{H}$, corresponding to different
realizations of the maps \cite{GladikowskiHopf1997}.

The Hopf number $Q_{H}$ also has an elementary geometric interpretation. The
\emph{preimage} of every point of the target space $S^{2}$ is isomorphic to
a circle. These circles are all linked, meaning that any given circle
intersects the disk spanned by any other circle. Thus, the Hopf number
represents the linkage number, which quantifies the degree of the linkage
between any two arbitrary circles \cite%
{GladikowskiHopf1997,Uedaknotsspinor,lyuHHGhopfions2024}.

\begin{figure*}[tbp]
{\includegraphics[width=6in]{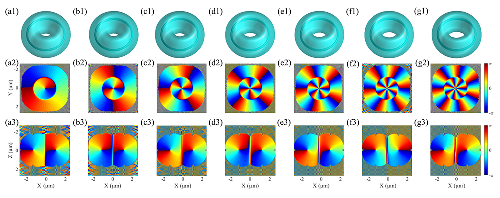}}
\caption{Stable hopfions with $S=1$, $M=1$ (a1-a3), $2$ \ (b1-b3), $3$
(c1-c3), $4$ (d1-d3), $5$ (e1-e3), $6$ (f1-f3), $7$ (g1-g3). The first row
shows the density isosurface. The second and third rows display the phase of
the wave function in the planes $Z=0$ and $Y=0$, respectively. The physical
parameters are set as $a=100$ $a_{0}$, $a^{\prime }=-110$ $a_{0}$, $R=0.9$ $%
\mathrm{\protect\mu }m$, $\Omega =51000$ Hz, and $N=111130$ (a1-a3), $109570$
(b1-b3), $107240~$(c1-c3), $104270$ (d1-d3), $100730$ (e1-e3), $96706$
(f1-f3), $92267$ (f1-f3).}
\label{hopfionsQHne0}
\end{figure*}

For the hopfions with $Q_{H}=0$, which we address in this section, different
circles corresponding to different points in the parameter space $S^{2}$ do
not link, resulting in concentric circles. To illustrate this property, Fig. %
\ref{equalvalueline} shows contour plots of different hopfions with $%
R_{0}=0.9~\mathrm{\mu }$m and $1.1\,\mathrm{\mu }$m. Solid circles in the
figure represent contours where both the real and imaginary parts of wave
function $\phi $ are constant (i.e., $\text{Re}(\phi )=C_{1}$, $\text{Im}%
(\phi )=C_{2}$), different colors corresponding to different values of $%
C_{1} $ and $C_{2}$. It is clearly observed that these circles are
concentric and do not intersect, which is consistent with $Q_{H}=0$. The
real and imaginary parts of the wave function represent two degrees of
freedom in the parameter space $S^{2}$, with different values corresponding
to different points on the $S^{2}$ sphere. Solid circles in the figure
represent the projection of a point from the $S^{2}$ sphere onto the real
space. The projection is realized through the Hopf map and stereographic
projection.

\subsection{Hopfions with $Q_{H}\neq 0$}

Hopfions with nonzero Hopf numbers are topologically nontrivial states.
Examples of hopfions with $S=1,M=1\sim 7$ are shown in Fig. \ref%
{hopfionsQHne0}. Accordingly, they carry $Q_{H}=1\sim 7$. The first row of
Fig. \ref{hopfionsQHne0} illustrates density isosurfaces of the hopfions,
revealing that they all are toroidal modes. The second and third rows
display the phase distributions of the wave functions in the horizontal ($%
Z=0 $) and vertical ($Y=0$) planes, respectively. The latter rows
demonstrate that the hopfions indeed have the twist number $S=1$, while
their vorticities in the horizontal plane range from $M=1$ to $M=7$. The
stability of these hopfions was verified by simulation of their perturbed
evolution, in the framework of Eq. (\ref{dimensionless-GPE}).

\begin{figure*}[tbp]
{\includegraphics[width=6in]{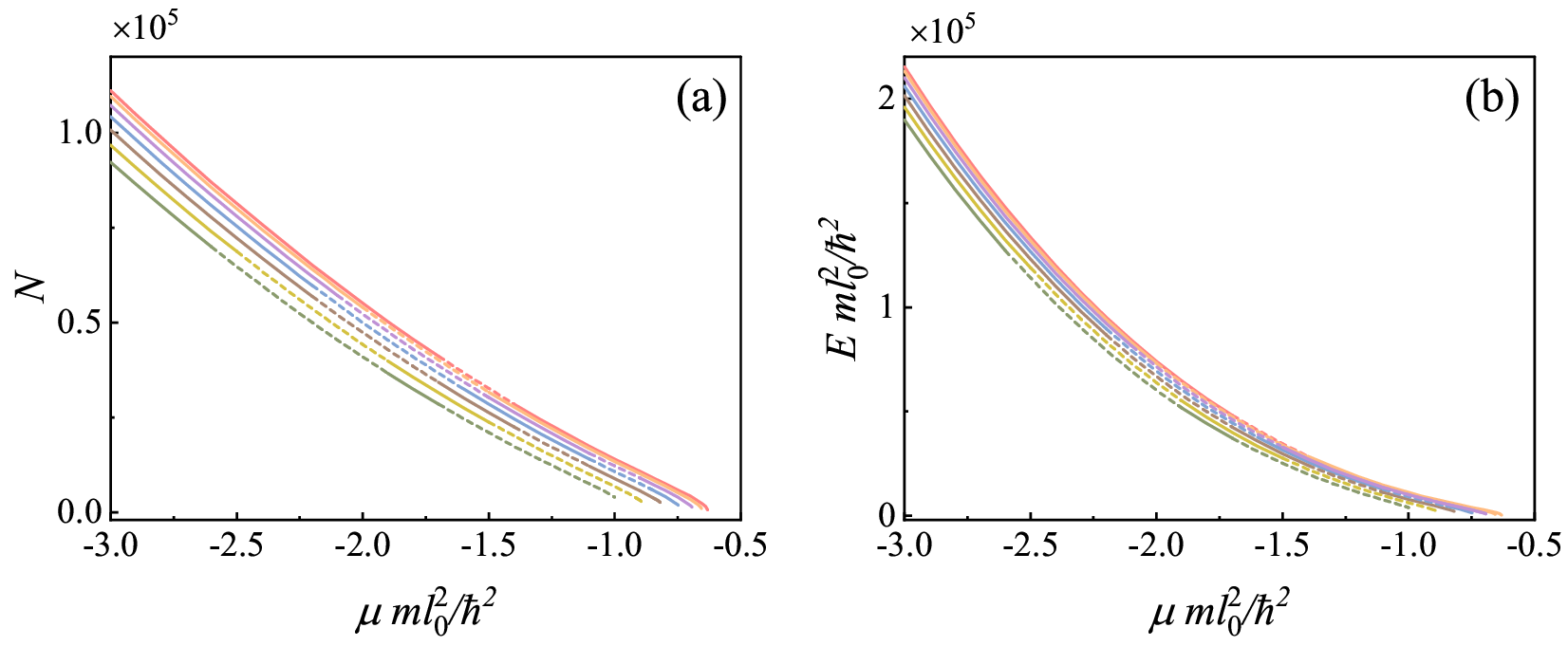}}
\caption{The atom number $N$ and energy $E$ vs. chemical potential $\protect%
\mu $ for the hopfions with $Q_{H}=1\sim 7$. Solid and dashed segments
denote stable and unstable solutions, respectively. The curves, ordered from
the top to bottom, correspond to values of the twist topological charge $%
M=1\sim 7$ in the increasing order, with fixed vorticity $S=1$. The physical
parameters are set as $a=100~a_{0}$, $a^{\prime }=-110~a_{0}$, $\Omega
=51000~$Hz, $R=0.9$ $\mathrm{\protect\mu }$m.}
\label{QHne0NEmu}
\end{figure*}

The $N(\mu )$ and $E(\mu )$ dependences for hopfions with $Q_{H}=1\sim 7$
are shown in Fig. \ref{QHne0NEmu}, where solid and dashed segments again
represent stable and unstable solutions, respectively. It is observed that,
as the twist winding number $M$ increases, the atom number $N$ and energy $E$
of the hopfions decrease. Similar to the above results for $Q_{H}=0$, the
hopfions with $Q_{H}=1\sim 7$ also exhibit a minimum-$N$ threshold, which
increases with the value of $Q_{H}$. Hopfions with higher Hopf numbers $Q_{H}
$ display intermittent instability regions. In general, the instability
region becomes broader as the Hopf number increases.

\begin{figure}[tbp]
{\includegraphics[width=3.2in]{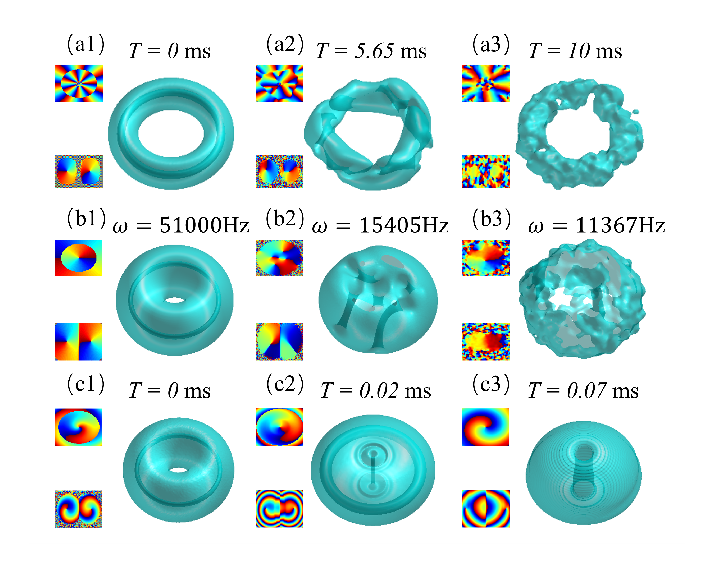}}
\caption{The unstable evolution of hopfions with $Q_{H}\neq 0$. (a1)-(a3): A
perturbed hopfion with $S=1$, $M=7$, $R_{0}=0.9~\mathrm{\protect\mu }\text{m,%
}$ $N=7523$, $a=100~a_{0}$, and $a^{\prime }=-110~a_{0}$. The evolution of
the hopfions with $S=1$, $M=1$, and $\Omega =51000\,\text{Hz}$ under the
action of a gradually decreasing TP strength $\Omega $ is shown in
(b1)-(b3). Panels (c1)-(c3) illustrate the evolution of a hopfion after the
TP was removed at the start of the evolution, with parameters $S=1$, $M=1$, $%
R_{0}=0.9~\mathrm{\protect\mu }\text{m}$, $N=111130$, $a=100~a_{0}$, and $%
a^{\prime }=-110~a_{0}$.}
\label{unstable}
\end{figure}

The evolution of unstable hopfions is illustrated in Fig. \ref{unstable}
(a1-a3). It is observed that the unstable hopfions lose their topological
structure in the course of the evolution. The vortex ring in the center is
gradually destroyed, and the hopfion eventually degenerates into the GS.
Similar to the unstable evolution reported in Ref. \cite{Bidasyukhopfsoliton}%
, the vortex ring in the center splits into fragments of vortex lines.
Additionally, we examined the evolution of a stable hopfion with parameters $%
S=1$, $M=1$, $\Omega =51000\,\mathrm{Hz}$, and $N=111130$. As shown in Fig. %
\ref{unstable} (b1-b3), the hopfion becomes unstable when the TP strength
falls below the critical value, $\Omega _{c}=15405\,\mathrm{Hz}$.
Furthermore, as displayed in Fig. \ref{unstable} (c1-c3), if the TP is
removed at the start of the evolution, the previously stable hopfion rapidly
becomes unstable on a short timescale $\simeq 0.02\ $ms. This observation
suggests that the hopfions, characterized by their complex topological
structure, are unlikely to remain stable in the free space, without the
support of the TP.

\begin{figure}[tbp]
{\includegraphics[width=3.2in]{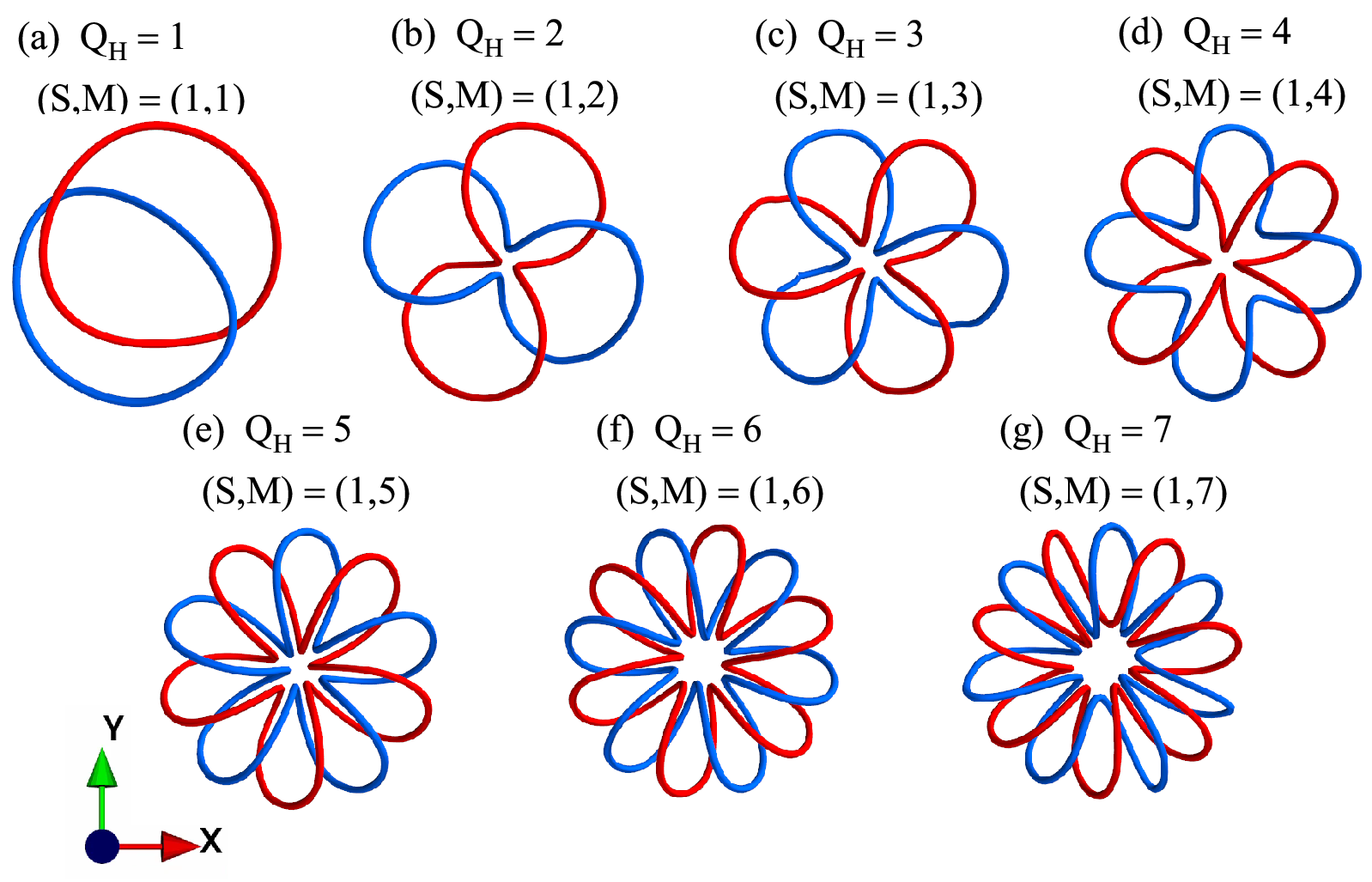}}
\caption{The elementary geometric interpretation of the Hopf numbers $%
Q_{H}=1 $ (a), $2$ (b), $3$ (c), $4$ (d), $5$ (e), $6$ (f), and $7$ (g) in
the horizontal plane, $Z=0$. The red and blue solid curves represent
distinct constant values of $\left( \mathrm{Re}(\protect\phi ),\mathrm{Im}(%
\protect\phi )\right) $: for $Q_{H}=1$, these are $(2,2)$ (red) and $(-2,-2)$
(blue); for $Q_{H}=2$, these are $(0.3,0.4)$ (red) and $(-0.3,-0.4)$ (blue);
for $Q_{H}=3$, these are $(-0.3,0.4)$ (red) and $(0.3,-0.4)$ (blue); for $%
Q_{H}=4$, these are $(-0.4,-0.3)$ (red) and $(0.4,0.3)$ (blue); for $Q_{H}=5$%
, these are $(-0.3,-0.2)$ (red) and $(0.2,-0.3)$ (blue); for $Q_{H}=6$,
these are $(-0.3,-0.3)$ (red) and $(0.3,0.3)$ (blue); and for $Q_{H}=7$,
these are $(-0.4,-0.4)$ (red) and $(0.4,0.4)$ (blue).}
\label{knots}
\end{figure}

Unlike the previously discussed case of $Q_{H}=0$, where the curves
corresponding to constant values of $(\mathrm{Re}(\phi ),\mathrm{Im}(\phi ))$
in the 3D space do not intersect, the hopfions with $Q_{H}\neq 0$ exhibit
intersection of these curves, resulting in a knot with a linking number
equal to $Q_{H}$. Figure \ref{knots} offers the elementary geometric
representations of the hopfions with $Q_{H}=1\sim 7$. In the figure, red and
blue curves correspond to distinct constant values of $(\mathrm{Re}(\phi ),%
\mathrm{Im}(\phi ))$. To better highlight the structure of the hopfions, the
figure focuses on the shape of the knots in the horizontal plane, $Z=0$,
where the linking number can be identified by observing which curve passes
over which one. Note that these are 3D curves, but not flat loops. When
viewed in the $Z=0$ cross-section, these knots exhibit a petal-like
structure, with the number of petals increasing with the Hopf number, $Q_{H}$%
.

\begin{figure}[tbp]
{\includegraphics[width=3.4in]{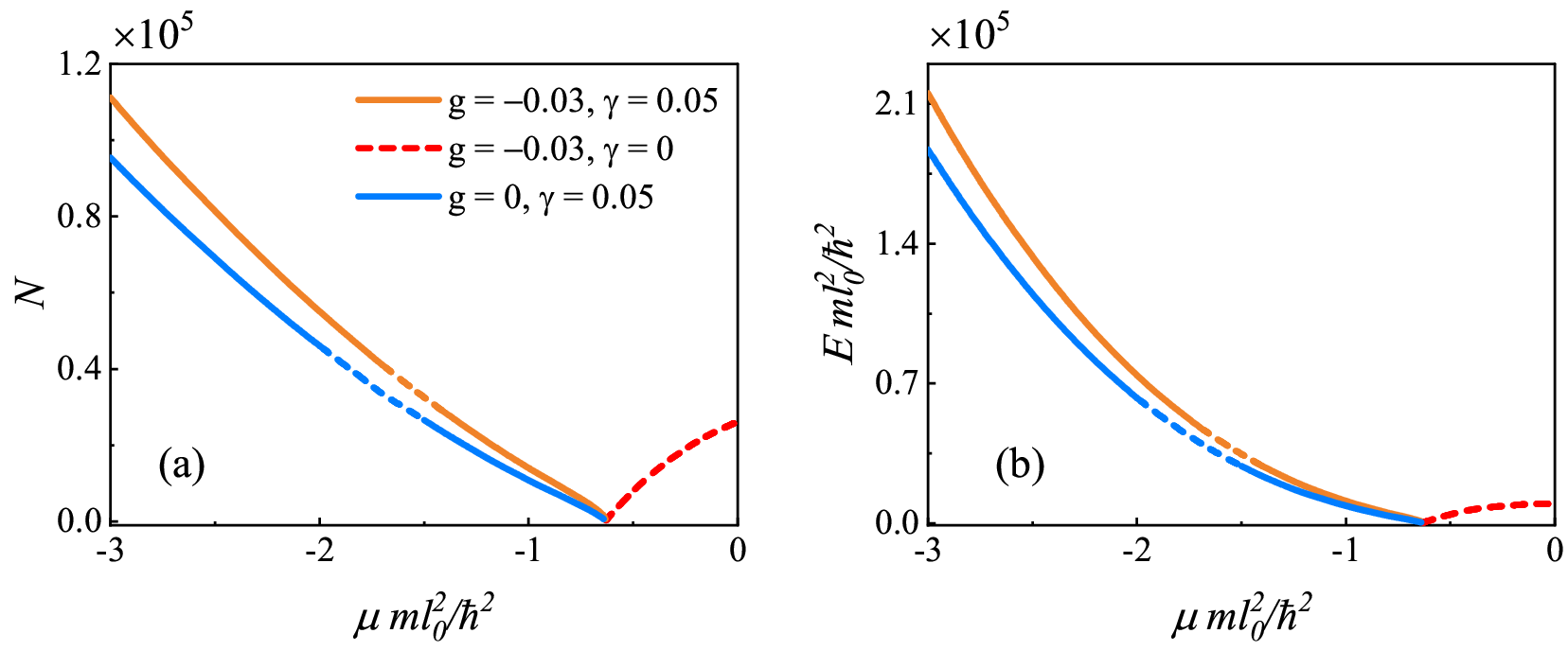}}
\caption{The $N(\protect\mu )$ (a) and $E(\protect\mu )$ (b) curves for
hopfion families with $S=1,M=1$. The orange curves represent an example of
the generic case considered above, with $g=-0.03,\protect\gamma =0.05$. The
red curves represent a fully unstable family in the absence of the LHY term,
with $\protect\gamma =0$ and $g=-0.03$. The blue curves display a partly
stable family found in the absence of the mean-field self-attraction, with $%
g=0\ $and $\protect\gamma =0.05$ (the \textquotedblleft LHY superfluid").
Solid and dashed segments correspond to stable and unstable families,
respectively.}
\label{appendix_NEmu}
\end{figure}

It is natural to expect that hopfions with $Q_{H}\neq 0$ cannot be stable in
the absence of the LHY correction, i.e., setting $\gamma =0$ in Eq. (\ref%
{dimensionless-GPE}). Indeed, our numerical analysis readily demonstrates
that, while the combination of the mean-field self-attractive cubic
nonlinearity and TP trap creates hopfions as solutions of Eq. (\ref%
{dimensionless-GPE}), they are completely unstable. In particular, in the
case of $\gamma =0$ the dependence $N(\mu )$ for the hopfion family with $%
S=1,M=1$ ($Q_{H}=1$), displayed in Fig. \ref{appendix_NEmu}(a) fully
contradicts the VK criterion, which confirms its complete instability.

\begin{figure}[tbp]
{\includegraphics[width=3.0in]{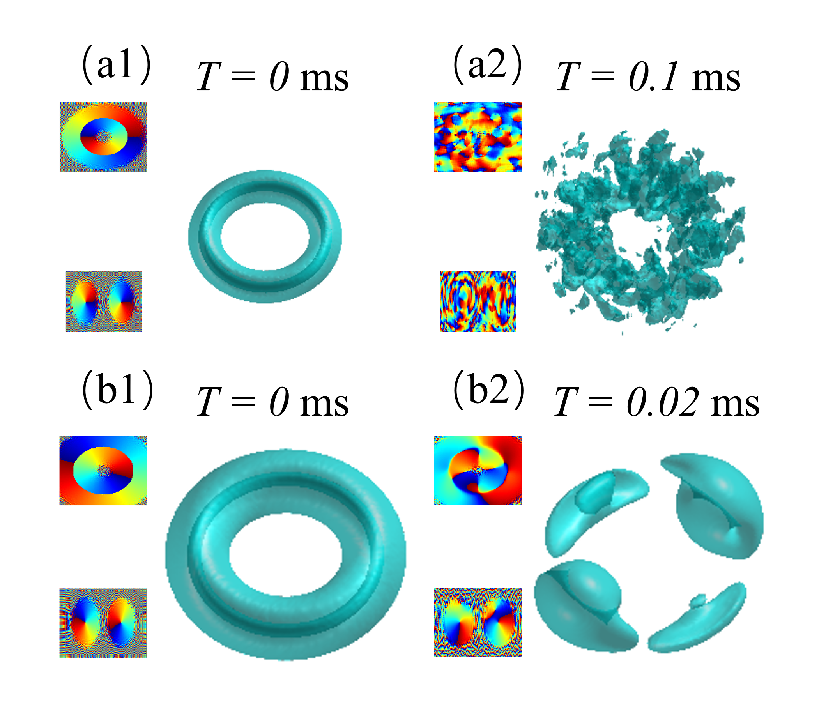}}
\caption{The unstable evolution of hopfions with $S=1,M=1$ in the absence of
the LHY term, \textit{viz}., $\protect\gamma =0$ and $g=-0.03$. Panels
(a1-a2) and (b1-b2) correspond, respectively, to values of the total norm $%
N=25.888$ and $N=223$.}
\label{unstable-mean}
\end{figure}

The unstable evolution of the hopfions in this case is illustrated in Fig. %
\ref{unstable-mean}. It is seen that the unstable hopfion with a relatively
small norm suffers complete destruction (in panel (a)), while a hopfion with
a large norm splits into a set of fragments, in panel (b).

Additionally, we examined the special case in which the nonlinearity is
represented solely by the LHY term, i.e., $g=0$ in Eq. (\ref%
{dimensionless-GPE}) (the so-called LHY superfluid \cite%
{jorgensenQuantumfluctation,LHY2}). In this case, we find that hopfion
families with $Q_{H}=1$ obey the VK criterion and are chiefly stable,
including an unstable subfamily, as demonstrated by the blue curves in Fig. %
\ref{appendix_NEmu}. The figure also shows that, for the same value of
strength $\gamma $ of the LHY term and $g=0$, the hopfions exhibit lower
norm $N$ and energy $E$, in comparison to their counterparts found in the
presence of the mean-field self-attraction ($g<0$). However, the instability
region is slightly larger in the case of $g=0$ than in the presence of $g<0$.

\section{Conclusion}

The objective of the work is to construct stable QDs (quantum droplets)
carrying the highly nontrivial topological structure which makes it possible
to identify them as hopfions. The stability of (a part of) these states,
which is a crucially important issue, is provided by the action of TP
(toroidal potential), which is included in the corresponding system of the
3D GPEs (Gross-Pitaevskii equations). The interaction in the GPE\ system is
represented by the self-attractive mean-field cubic terms and repulsive LHY
(Lee-Huang-Yang) quartic ones, which represent the effect of quantum
fluctuations around the mean-field configurations. The numerical analysis
demonstrates that not only the modes with zero Hopf number, $Q_{H}=0$ (they
have zero vorticity, $M=0$, and the twist topological charge $S=1$) can
stably exist in this system, but also the hopfions in the range of $%
Q_{H}\equiv MS=1\sim 7$ exhibit stability regions. The intensity profiles of
these solutions in the horizontal ($Z=0$) plane reveal a characteristic
double-ring \textit{preimage} of the hopfion, which may serve as a distinct
signature for identifying hopfions in the experiment. Naturally, the
stability regions gradually shrink with the increase of $Q_{H}$. In the
course of their evolution, unstable hopfions lose their topological
structures, eventually decaying into the GS (ground state). The hopfions
cannot be stable if the confining TP is absent, i.e., they cannot represent
stable states in the free space. To further elucidate the geometry and
topology of the hopfions (in particular, of the stable ones), we have
displayed their elementary geometric representation by means of the
corresponding Hopf map (of the parameter manifold into the real space) and
stereographic projection. For $Q_{H}=0$, the representation reveals a series
of non-intersecting concentric rings in the real space. In contrast, the
solutions with $Q_{H}=1\sim 7$ exhibit intricate knots, with the linking
number equal to $Q_{H}$. In the horizontal plane, these knots exhibit a
petal-like structure, with the number of petals also equal to $Q_{H}$.
Furthermore, the role of the LHY term is crucial in determining the 
stability of Hopfions. When the LHY term is absent (\(g<0,\gamma = 0\) in 
Eq. (\ref{dimensionless-GPE})), 
Hopfions with \(Q_H \neq 0\) cannot stably exist. In contrast, when 
only the LHY nonlinear term is present (\(g = 0, \gamma \neq 0\) in 
Eq. (\ref{dimensionless-GPE})), 
Hopfions with \(Q_H \neq 0\) can stably exist, although the instability 
region is slightly larger compared to the case with \(g < 0\).

\begin{acknowledgments}
This work was supported by NNSFC (China) through Grants No. 12274077,
12475014, the Natural Science Foundation of Guangdong
province through Grant No. 2024A1515030131, 2025A1515011128, 2023A1515010770,
Guangdong Basic and Applied Basic Research Foundation Grant No. 2023A1515110198,
the Research
Fund of Guangdong-Hong Kong-Macao Joint Laboratory for Intelligent
Micro-Nano Optoelectronic Technology through grant No.2020B1212030010. The
work of B.A.M. is supported, in part, by the Israel Science Foundation
through grant No. 1695/2022.
\end{acknowledgments}

\end{document}